\documentclass[runningheads]{svmult}

\usepackage{makeidx}   
\usepackage{graphicx}  
\usepackage{subeqnar}  
\usepackage{multicol}  
\usepackage{physprbb}  
\makeindex             

\newcommand{\smean}[1]{\langle #1 \rangle} 

\begin{document}
\title*{Capacitance, Charge Fluctuations 
and Dephasing In Coulomb Coupled Conductors}
\toctitle{Capacitance, Charge Fluctuations and
\protect\newline Dephasing In Coulomb Coupled Conductors}
%
%
\titlerunning{Capacitance, Charge Fluctuations }
%
\author{Markus {B\"{u}ttiker}}
\authorrunning{Markus B\"{u}ttiker}
%
%
\institute{Universit\'e de Gen\`eve,
D\'epartement de Physique Th\'eorique,\\
CH-1211 Gen\`eve 4, Switzerland}

\maketitle              

\begin{abstract}
The charge fluctuations of two nearby mesoscopic conductors
coupled only via the long range Coulomb force are discussed and used 
to find the dephasing rate which one conductor exerts on the other. 
The discussion is based on a formulation 
of the scattering approach for charge densities and the density response 
to a fluctuating potential. Coupling to the Poisson equation 
results in an electrically self-consistent description of charge 
fluctuations. At equilibrium 
the low-frequency noise power can be expressed with the help 
of a charge relaxation resistance (which together with the capacitance 
determines the RC-time of the structure). In the presence of transport 
the low frequency charge noise power is determined by a resistance 
which reflects the presence of shot noise. 
We use these results to derive expressions for the 
dephasing rates of Coulomb coupled conductors and to find  
a self-consistent expression for the measurement time. 
\end{abstract}

\section{Introduction}

\begin{figure}
\begin{center}
\includegraphics[scale=1.2]{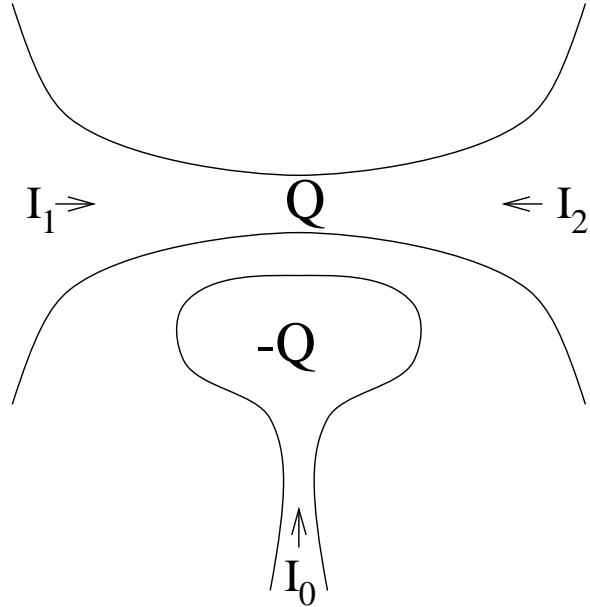}
\end{center}
\caption{Cavity with a charge deficit $-Q$ in the proximity of a quantum point contact
with an excess charge $Q$. The dipolar nature of the charge distribution 
ensures the conservation of currents $I_{0}$, $I_{1}$ and $I_{2}$ 
flowing into this structure. After \protect\cite{plb} . 
}
\label{ped}
\end{figure}

Investigations of time-dependent current fluctuations 
of mesoscopic systems have been widely used 
to obtain information 
which cannot be extracted from conductance measurements 
alone \cite{physr}. In this work we are interested 
in the fluctuations of the {\it charge} in a volume element inside 
the electrical conductor. If the volume element is made very small 
the fluctuations of interest are thus the fluctuations 
of the {\it local} electron density. Such fluctuations can be detected, 
for instance by measuring the current induced into a nearby
gate \cite{plb} or a small cavity as shown in Fig. \ref{ped}. 
Through the long range Coulomb interaction a 
charge fluctuation above the average equilibrium value of the charge 
in the conductor 
generates additional electric fields which lead to a charge reduction 
at the surface of the gate. The reduction of charge at the gate surface is 
accomplished by a flow of carriers out of the contact of 
the gate \cite{plb}. The conductor can be in an equilibrium state 
in which case the charge fluctuations are associated with 
Nyquist noise or it can be in a transport state and the charge 
fluctuations are those that are generated by shot noise. 

Charge fluctuations can be detected not only by direct capacitive 
probing.
In an experiment by Buks et al. \cite{buks1}
charge fluctuations are observed 
through conductance measurements: the charge fluctuations 
of two conductors in close proximity can give rise to 
an additional dephasing rate which a carrier in one 
conductor experiences due to the presence of the 
other conductor. In the experiment 
of Buks et al. \cite{buks1}  an Aharonov-Bohm ring with a quantum dot 
in one of its arms is in close proximity to another conductor
which forms a QPC (quantum point contact). The presence of the 
QPC leads to a reduction of
the Aharonov-Bohm interference oscillations. In the experiment of 
Sprinzak et al. \cite{buks2} a double quantum dot is brought 
into the proximity of a conductor in a high magnetic field 
and the broadening of the Coulomb blockade peak due the charge fluctuations 
in the edge states of the nearby conductor are measured. 
Theoretical discussions of dephasing rates in coupled systems
are given in 
Refs. \cite{harris,gurvitz,aleiner1,levinson,stod,mbam,sitges}. 
Harris and Stodolsky \cite{harris,stod} view this as a quantum measurement 
problem in which the state of one system is measured 
with the help of another. In this work, as well as in the experimental work 
of Buks et al. \cite{buks1} the time it takes to ascertain the state of the 
measured system, the measurement time, is identified with the dephasing 
time. 
Gurvitz investigates the time-evolution 
of the density matrix of the system that is measured \cite{gurvitz}. 
Aleiner et al. \cite{aleiner1} relate the dephasing rate to the 
orthogonality catastrophe which occurs if an additional carrier is added 
to the ground state of the system. Levinson derives a dephasing rate 
in terms of the charge fluctuation spectrum of 
non-interacting carriers \cite{levinson}. 
The approach which we discuss here also 
relates the dephasing rate to the charge fluctuation spectrum. 
However, in contrast to the discussions presented in 
Refs.  \cite{harris,gurvitz,levinson,stod} we emphasize an electrically 
self-consistent approach which takes into account that charging 
even in open conductors such as QPC can be energetically 
expensive \cite{mbam}. 
This approach is applicable to a wide range of geometries 
and in Refs. \cite{mbam,sitges} has been used to present a 
self-consistent treatment of charge fluctuations 
in edge states. Our discussion can be compared with Ref. \cite{levins2}
which treats fluctuations as a free electron problem.

It is interesting to notice that 
in discussions of charge fluctuations in systems 
that are composed entirely of components in which charge quantization is 
important, theoretical work  \cite{ss,mss,ak,averin,korot}
carefully discusses the various capacitance 
coefficients which determine the charging energies of the 
system and the coupling. On the other hand, when subsystems like 
QPC's are discussed, electric carriers are treated 
as if they were non-interacting 
entities \cite{mss,ak,averin,korot}. Clearly, a QPC
has also a capacitance. A QPC can be characterized 
by a capacitance which describes its self-polarization \cite{chris,guo,wang}. 
This self-polarization corresponds to charge accumulation 
on one side of the quantum point contact and charge depletion on the other 
side of the quantum point contact \cite{chris,guo}. 
The self-polarization does not change the overal charge of the QPC
and is thus here not of primary importance. But a 
quantum point contact can be charged vis-a-vis the gates \cite{sohn}
or vis-a-vis any other conductor. This leads to a net charge on the quantum 
point contact and is the dominant process by which the quantum point
contact interacts with an other conductor \cite{sohn}. 
 
Therefore, like in the systems which exhibit Coulomb blockade, 
we can ask: "What is the dependence of a charge fluctuation spectrum of 
a QPC on its capacitance?", or "How does the additional 
dephasing rate generated by the proximity of a QPC 
depend on the capacitance of the quantum point contact?" 
If the relevant capacitance were a mere geometrical quantity 
these questions might just determine some prefactors left open 
in previous work. However, the charge screening of a QPC 
is non-trivial since the density of states of a quantum point contact \cite{plb} 
(at least in the semi-classical limit) diverges for a gate voltage 
at which a new channel is opened (quantum tunneling limits the 
density of states). 

The aim of this work is to present a simple self-consistent discussion 
of charge fluctuation spectra of Coulomb coupled conductors and 
to use these spectra to find dephasing rates
and self-consistent expression for the measurement time. 
We assume that the ground state of the system has been determined 
and investigate small deviations away from the ground state.
The approach 
which we present combines the scattering approach with the Poisson equation 
and treats interactions in the random phase approximation \cite{mbam}. 
This approach 
has been used with some success to treat the dynamic conductance 
\cite{btp,math}
of mesoscopic systems, their non-linear I-V-characteristics \cite{chris2,ma,kir}, 
and higher harmonics generation \cite{ma}.  
Indeed there is a close connection between the charge fluctuation 
spectra which we obtain and the dynamic conductances of a mesoscopic
system \cite{btp,sohn}. We show that the charge noise power of the equilibrium 
charge fluctuations can at low frequencies be characterized by 
a charge relaxation resistance $R_q$. Together with the capacitance 
this resistance determines the $RC$-time of the mesoscopic 
structure \cite{btp}.  
Indeed, independent measurements of 
the capacitance and the resistance $R_q$, when compared 
with the results from a measurement of the dephasing 
rate would provide an important 
overall test of the consistency of the theory. 
In the presence of transport, we deal in the low temperature limit 
with shot noise. In this case, 
the low frequency charge noise power is proportional 
to the applied voltage and proportional to 
a resistance $R_v$. 

Some works advocate a perturbative treatment
of coupled systems and present results which are proportional 
to the square of the coupling constant. In contrast, the self-consistent 
approach discussed here 
leads to a more intersting dependence on 
the bare coupling constant \cite{mbam} (in the physically most 
relevant case we actually find a dephasing 
rate which is proportional to 
one over the square of the coupling strength).

In this work, we treat only the symmetric QPC 
Coulomb coupled to another system. This limitation is motivated by the 
fact that the QPC is the most widely investigated example. 
This should permit a most direct comparison of 
the self-consistent approach 
advocated here with the results in the literature. For a review which addresses
a wider range of geometries we refer the reader to \cite{ankara}. 
We also restrict ourselves to the case where linear screening is applicable.

An evaluation of the dephasing rates requires a discussion of the 
potential fluctuations (or equivalently,
the charge fluctuations). We will need 
only the zero-frequency limit of the potential fluctuation spectrum, 
and it is thus sufficient to find the zero-frequency, white noise limit
of the charge fluctuations. 

\section{The effective interaction} 

To be definite, consider the 
conductors of Fig.~\ref{ped}. We first investigate the relation 
between voltage and charges of these two conductors 
for small deviations of the applied voltages away from 
their equilibrium value. 
We assume that each electric field line emanating from 
the cavity ends up either again on the cavity or on the QPC. 
There exists a Gauss volume which encloses both conductors
which can be chosen large enough so that the electrical flux
through its surface vanishes \cite{math}. Consequently, 
the charge on the two conductors is conserved. 
Any accumulation of charge at one location 
within our Gauss volume is compensated by a charge depletion 
at another location within the Gauss volume. 
The variation of the charge brought about by a small 
change of the applied voltage is thus of a dipolar nature. 

To keep the discussion simple, we consider here the case that each conductor 
is described by a single potential only. The charge and potential 
on the cavity are denoted by $dQ_{0}$ and $dU_{0}$ and on the 
QPC by $dQ_{1}$ and $dU_{1}$. 
A more accurate description can be obtained by subdividing the conductor 
into a number of volume elements \cite{sohn} or in fact by using 
a continuum description \cite{math}. 
The essential elements of our discussion do, however, already become 
apparent in the simple case that each conductor is described only by one
potential and we will treat only this limiting case in this work. 
The Coulomb interaction is described with the help 
of a single {\it geometrical} capacitance $C$. 
The charges and potentials of the two conductors are then related by
\begin{equation}
d{Q}_{0} = C(d{U}_{0} - d{U}_{1}),
\label{q1}
\end{equation}
\begin{equation}
d{Q}_{1} = C(d{U}_{1} - d{U}_{0}).
\label{q2}
\end{equation} 
The two equations can be thought off as a discretized version 
of the Poisson equation. Note that according to Eqs. (\ref{q1},\ref{q2})
we have $dQ_{0} + dQ_{1} = 0 $. 

We now complement these two equations by writing the charges 
$dQ_{i}$ as a sum of an external (or bare) charge calculated 
for fixed internal potentials $U_{0}$ and $U_{1}$ and 
an induced charge generated by the response of the potential 
due to the injected charges. The additional charge injected into the 
cavity (which we denote by $e dN_{0}$) 
due to an increase of the reservoir voltage by $edV_{0}$ is 
$e dN_{0} = eD_{0}edV_{0}$. Here $D_{0}$ is the total density of states at the 
Fermi energy of the cavity of the region in which $dU_{0}$ deviates 
from its equilibrium value. The induced charge is $- e^{2} D_{0} dU_{0}$. 
It is negative since the Coulomb interaction counteracts charging. 
It is also determined by the total density of states since the integrated 
Lindhard function is given by the density of states \cite{note1}. 
Thus the charge on the cavity is 
\begin{equation}
d{Q}_{0} = (edN_{0} - e^{2} D_{0} dU_{0}).
\label{q3}
\end{equation} 
To find the charge on the QPC we must take into account 
that there are two reservoir potentials which we denote by $dV_{1}$
and $dV_{2}$. An increase in the potential of contact of reservoir $1$
(at constant internal potential $U_{1}$) 
does not fill all the available states but only a 
portion. The density of states \cite{sohn} of carriers incident from reservoir 
$1$ is denoted by $D_{11}$
and is called the {\it injectance} 
of contact $1$. Thus the charge injected at constant internal 
potential by an increase of the voltage $V_{1}$ is $eD_{11}edV_{1}$.
Similarly an increase of the potential at reservoir $2$
leads to an additional charge 
$eD_{12}edV_{2}$.  Here $D_{12}$
is the injectance of reservoir $2$. Together these two 
(partial) density of states are equal to the total density of states 
of the quantum point contact $D_{11} + D_{12} = D_{1}$. 
Thus the total injected charge on conductor $1$ 
is $edN_{1} = eD_{11}edV_{1} + eD_{12}edV_{2}$. Screening is against 
the total density of states and thus a variation of the internal 
potential $dU_{1}$ generates an induced charge 
given by $e^{2}D_{1}dU_{1}$. The charge $dQ_{1}$ is the sum of these 
three contributions, 
\begin{equation}
d{Q}_{1} = (edN_{1} - e^{2}D_{1} dU_{1}).
\label{q4}
\end{equation}
We arrive thus at the following self-consistent 
equations relating charges and potentials, 
\begin{equation}
d{Q}_{0} = C(d{U}_{0} - d{U}_{1}) = (edN_{0} - e^{2} D_{0} dU_{0}),
\label{q5}
\end{equation} 
\begin{equation}
d{Q}_{1} = C(d{U}_{1} - d{U}_{0}) = (edN_{1} - e^{2} D_{1} dU_{1}). 
\label{q6}
\end{equation} 
We can use these equations to express the internal potentials
$U_{i}$ in terms of the injected charges $edN_{i}$.
We find $dU_{i} = e \sum_j G_{ij} {dN}_{j} $
with an 
effective interaction $G_{ij}$ given by 
\begin{equation}
    {\bf G} = \frac{C_{\mu}}{e^{2}D_{0}e^{2}D_{1}C} \left( \begin{array}{ll}

        C + e^{2}D_{1} & C \\

        C &  C + e^{2}D_{0}

        \end{array} \right)
        \label{geff} .
\end{equation}
Here $C_{\mu}$ is 
the electrochemical capacitance 
\begin{equation}
C_{\mu}^{-1} = C^{-1} + (e^{2}D_{0})^{-1}+  (e^{2}D_{1})^{-1} 
\label{cmu}
\end{equation} 
which is 
the series capacitance of the geometrical contribution $C$ and the 
density of states of the two conductors \cite{btp}. 
Note that in contrast to perturbation treatments, 
the effective coupling element $G_{12}$ is not proportional to $e^{2}/C$
but in general is a complicated function of this energy. 
We will use the effective interaction in Section $7$
to find the measurement time. First, however, we will now use 
the effective interaction to express the true charge fluctuations 
in terms of the bare fluctuations.

We are interested not in the average quantities discussed above 
but in their dynamic fluctuations. To this extend we now re-write 
Eqs. (\ref{q5}) and (\ref{q6}) for the fluctuating quantities. 
In a second quantization approach the fluctuating quantities are described 
with the help of operators, ${\hat Q}_{i}$ for the true charges, 
and the potentials ${\hat U}_{i}$ on the two conductors, $i =0,1$. 
As for the average charge, the fluctuating charge can also 
be written in terms of bare charge fluctuations 
$e {\hat {\cal N}}_{i}$ 
(calculated by neglecting the Coulomb interaction) 
counteracted by a screening charge $e D_{i} e{\hat U}_{i}$.
Below, we will give explicit expressions for all these operators.
Instead of Eqs. (\ref{q5}) and (\ref{q6}) 
we now have, 
\begin{equation}
{\hat Q}_{0} = C({\hat U}_{0}-{\hat U}_{1}) = 
e {\hat {\cal N}}_{0} - e^{2} D_{0 } {\hat U}_{0},
\label{f1}
\end{equation} 
\begin{equation}
{\hat Q}_{1} = C({\hat U}_{1}-{\hat U}_{0}) =
e {\hat {\cal N}}_{1} - e^{2} D_{1} {\hat U}_{1} . 
\label{f2}
\end{equation} 
Clearly, if we consider simply the average of these equations,
they must reduce to Eq. (\ref{q5}) and (\ref{q6}). 
The fluctuations are determined by the off-diagonal 
elements of the charge and potential operators. 
Below we will specify these expressions in detail. 
Solving these equations for the potential operators, we find 
$\hat U_{i} = e \sum_j G_{ij} \hat{ \cal N}_{j} $
with the 
effective interaction $G_{ij}$ given by Eq. (\ref{geff}). 

Let us now introduce the noise power spectra of the bare 
charges, $S_{N_{i}N_{i}}(\omega)$ for each of the conductors.
The spectrum of the bare charge fluctuations 
is defined as 
\begin{equation}
S_{N_{i}N_{i}}(\omega)2\pi\delta(\omega+\omega')=
\smean{\hat{\cal N}_{j}(\omega) \hat{\cal N}_{j}(\omega')
+ \hat{ \cal N}_{j}(\omega') \hat{ \cal N}_{j}(\omega)}
\label{sdef}
\end{equation} 
with $\hat{ \cal N}_{j}(\omega)=
\hat{\cal N}_{j}(\omega)-\smean{\hat{ \cal N}_{j}(\omega)}$, where 
$\hat{\cal N}_{j}(\omega)$ is the Fourier transform of the charge 
operator of conductor $j$. 
 
The bare charge fluctuation spectra on different conductors are 
uncorrelated,  $S_{N_{i}N_{j}}(\omega) = 0$ for $i \ne j$.
With the help of the effective interaction matrix, we can 
now relate the potential fluctuation spectra to the fluctuation 
spectra of the bare charges. In the zero-frequency limit we find, 
\begin{equation}
S_{U_{i}U_{j}}(0) = e^{2} \sum_k G_{ik} G_{jk} S_{N_{k}N_{k}}(0) .
\label{ub}
\end{equation} 
Even though the bare charge fluctuations are uncorrelated, 
the potential fluctuations and the true charge fluctuations 
on the two conductors are correlated.

\section{Charge Relaxation Resistances} 

It is useful to characterize the noise power of 
the charge fluctuations with the help of resistances. 
Consider first the case where both conductors are at equilibrium. 
Charge fluctuations on the conductors arise due to the random thermal 
injection of carriers. 
The bare charge fluctuation spectrum, normalized 
by the density of states $D_{i}$ of conductor $i$ 
has the dimension of a resistance. We introduce the 
charge relaxation resistance $R_{q}^{(j)}$ of conductor $j$, 
\begin{equation}
2kT R_{q}^{(j)} \equiv e^{2}S_{N_{j}N_{j}}(0)/(e^{2}D_{j})^{2}.
\label{rq}
\end{equation} 
The charge relaxation resistance has a physical significance 
in a number of problems. 
In simple cases, $R_{q}$ together with an appropriate 
capacitance determines the $RC$-time of the mesoscopic 
structure \cite{btp}. 
The charge relaxation resistance can thus alternatively be determined 
by investigating the poles of the conductance matrix \cite{math,btp}. 
The dynamic conductance matrix $G_{\alpha\beta} (\omega) \equiv 
dI_{\alpha} (\omega)/dV_{\beta}(\omega)$
of our mesoscopic structure (QPC and cavity) 
which relates the currents $dI_{\alpha} (\omega)$ at a frequency $\omega$
at contact $\alpha$ to the voltages $dV_{\beta}(\omega)$ applied at contact 
$\beta$ has at low frequencies a pole determined by $\omega_{RC} = -i 
C_{\mu} (R_{q}^{(1)} + R_{q}^{(2)})$. Alternatively we could carry out 
a low frequency expansion of the element $G_{00}(\omega)$ (the element 
of the conductance matrix which gives the current at the contact of the cavity
in response to an oscillating voltage applied to the cavity)
to find \cite{btp,math} that 
$G_{00}(\omega) = -i C_{\mu} \omega + C^{2}_{\mu} R_{q} 
\omega^{2} +..$. Thus $R_q$ plays a role in 
many problems. The charge relaxation resistance differs 
from the dc-resistance. For instance a ballistic one-channel 
quantum wire connecting two reservoirs and capacitively coupled to a gate 
has for spinless carriers a dc-resistance of $R = h/e^{2}$ and 
a charge relaxation resistance \cite{bhb} of $R_{q} = h/4e^{2}$. 
The dc-resistance corresponds to the series addition of resistances along
the conductance path, whereas an excess charge on the conductor relaxes
via all possible conductance channels to the reservoirs and thus corresponds
to the addition of resistances in parallel. This is nicely illustrated 
for a chaotic cavity \cite{bb} connected via contacts with $M_{1}$ and $M_{2}$ perfectly
transmitting channels to reservoirs and capacitively coupled to a
gate. Its ensemble averaged
dc-resistance is $R = (h/e^{2})(M_{1}^{-1} + M_{2}^{-1})$, 
whereas its charge relaxation resistance is 
$R_{q} = (h/e^{2})(M_{1}+ M_{2})^{-1}$. Thus the dc-resistance 
is governed by the smaller of the two contacts, whereas the charge 
relaxation resistance is determined by the larger contact.

If the conductor is driven out of equilibrium
with the help of an applied voltage $|V| \equiv |V_{1}-V_{2}|$, the thermal 
noise described by Eq. (\ref{rq}) can be overpowered by shot noise. 
For $e|V| \gg kT$ 
the charge fluctuation spectrum becomes proportional to the applied voltage 
and defines a resistance $R_{v}^{(j)}$ via the relation, 
\begin{equation}
2e|V| R_{v}^{(j)} \equiv e^{2} S_{N_{j}N_{j}}(0)/(e^{2}D_{j})^{2}.
\label{rv}
\end{equation} 
The resistance $R_{v}$ is thus a measure of the 
noise power of the charge fluctuations associated with 
shot noise.   

Eq. (\ref{rq}) and Eq. (\ref{rv}) describe the behavior 
of $e^{2} S_{N_{j}N_{j}}(0)/(e^{2}D_{j})^{2}$
in the limits $kT \gg e|V|$ and $kT \ll e|V|$. For fixed temperature as 
a function of voltage $e^{2} S_{N_{j}N_{j}}(0)/(e^{2}D_{j})^{2}$
exhibits a smooth crossover from the equilibrium result Eq. (\ref{rq})
to Eq. (\ref{rv}) valid in the presence of shot noise. 
For the structure shown in Fig. \ref{ped} 
it is only the QPC (conductor $1$) which can be brought out 
of equilibrium. The cavity, connected to a single lead 
always exhibits only thermal fluctuations and its 
charge relaxation resistance is characterized by $R_q^{0}$
even if the QPC is subject to shot noise.

\section{Bare Charge Fluctuations and the Scattering Matrix} 

Let us now determine the charge operator for the bare charges
(non-interacting carriers). 
The operator for the total charge on a mesoscopic 
conductor can be found from the current operator  
and by integrating the continuity 
equation over the total volume of the 
conductor. This gives a relation 
between the charge in the volume and the particle 
currents entering the volume.  
We obtain for the density operator \cite{plb} 
\begin{equation} \label{qop}
\hat{\cal N}(\omega) = {\hbar} \sum_{\beta\gamma}
\sum_{mn} \int dE \: \hat a^{\dagger}_{\beta m}(E) 
{\cal D}_{\beta\gamma mn} (E, E + \hbar \omega) 
\hat a_{\gamma n} (E + \hbar \omega),
\end{equation}
where ${\hat a}^{\dagger}_{\beta m}(E)$ 
(and  ${\hat a}_{\beta m}(E)$) creates (annihilates) an incoming particle 
with energy $E$ in lead 
$\beta$ and channel $m$. 
The element 
${\cal D}_{\beta\gamma mn} (E, E + \hbar \omega)$
is the non-diagonal density of states element 
generated 
by carriers incident simultaneously in contact $\beta$ in quantum channel $m$
and by carriers incident in contact $\gamma$ in channel $n$. 
In particular, in the zero-frequency limit, 
we find in matrix notation \cite{plb,mbam}, 
\begin{equation} \label{qop2}
{\bf {\cal D}}_{\beta\gamma} (E) =  \frac{1}{2\pi i}
\sum_{\alpha}  
{\bf s}^{\dagger}_{\alpha\beta}(E)
\frac{d{\bf s}_{\alpha\gamma} (E)}{dE} .
\end{equation}
Expressions of this type are known from the discussion of quantum mechanical 
time delay \cite{smith}. 
The sum of the diagonal elements 
of this matrix is the density of states of the conductor
\begin{equation} \label{den}
D (E) = \sum_{\beta} 
Tr[{\bf {\cal D}}_{\beta\beta} (E)] =  \frac{1}{2\pi i}
\sum_{\alpha,\beta}  
Tr[{\bf s}^{\dagger}_{\alpha\beta}(E)
\frac{d{\bf s}_{\alpha\beta} (E)}{dE}] ,
\end{equation}
where the trace is over the quantum channels. 
${\cal D}_{\beta\beta}\equiv {\cal D}_{\beta}$ is the injectance 
of contact $\beta$. (We used this density of states in the discussion leading to 
Eq. (\ref{q3})). 

The charge fluctuations 
are determined by the off-diagonal elements of Eq. (\ref{qop}). 
Proceeding as for the case of current fluctuations \cite{leso,mb92,physr}
we find for the fluctuation spectrum of the total charge 
\begin{eqnarray} \label{snn}
S_{NN} (0) = {2h}
\sum_{\gamma\delta} \int dE \:\,
Tr[{\cal D}^{\dagger}_{\gamma\delta}{\cal D}_{\delta\gamma}]
f_{\gamma} (E) (1 - f_{\delta} (E)) . 
\end{eqnarray}
The spectrum of the bare charge fluctuations has to be determined 
for each conductor separately using its scattering matrix. 
We now go on to find specific expression for this spectrum for 
the QPC.

\section{Charge relaxation resistance of a quantum point contact} 

To illustrate the preceding discussion, we now consider specifically 
the charge relaxation resistance and subsequently the 
resistance $R_{v}$ of a QPC. 
For simplicity, we consider a
symmetric QPC (the asymmetric case \cite{buks2}
is treated in \cite{mbam,levins2}): 
For a symmetric scattering potential  
the scattering matrix (in a basis in which the 
transmission and reflection matrices are diagonal) is 
for the $n$-th channel of the form
\begin{equation}
    s_{n}(E) = \left( \begin{array}{ll}
	-i \sqrt{R_{n}} \exp(i\phi_{n}) & \sqrt{T_{n}} \exp(i\phi_{n}) \\
	\sqrt{T_{n}} \exp(i\phi_{n}) & -i\sqrt{R_{n}} \exp(i\phi_{n})
	\end{array} \right) ,
	\label{sqpc}
\end{equation} 
where $T_{n}$ and $R_{n}= 1-T_{n}$
are the transmission and reflection probabilities 
and $\phi_{n}$ 
is the phase accumulated by a carrier in the $n$-th eigen channel. 
We find for the elements 
of the density of states matrix, Eq. (\ref{qop2}),
\begin{equation}
    {\cal D}_{11} = {\cal D}_{22} = 
    \frac{1}{2\pi} \frac{d\phi_{n}}{dE} ,\,\, 
    {\cal D}_{12} = {\cal D}_{21} = \frac{1}{4\pi}  
    \frac{1}{\sqrt{R_{n}T_{n}}} \frac{dT_{n}}{dE} .
\label{mqpc}
\end{equation}
With these density of states matrix elements, we can determine the 
particle fluctuation spectrum, Eq. (\ref{snn})
in the white-noise limit, and  $R_q$ with the help of 
Eq. (\ref{rq})
\begin{equation}
R_{q} = \frac{h}{4e^{2}} 
\frac{\sum_{n} \left[(\frac{d\phi_{n}}{dE})^{2} + 
\frac{1}{4T_{n}R_{n}} (\frac{dT_{n}}{dE})^{2} \right]}
{[\sum_{n} \frac{d\phi_{n}}{dE}]^{2}} .
\label{rqgen}
\end{equation}
Eq. (\ref{rqgen}) is still a formal result, applicable 
to any symmetric (two-terminal) conductor. 
To proceed we have to adopt a specific model 
for a QPC. 
If only a few channels are open the average potential 
has in the center of the conduction channel 
the form of a saddle \cite{mbqpc}:
\begin{equation}
    V_{eq}(x,y) = V_0 + \frac{1}{2} m \omega_y^2 y^2
    - \frac{1}{2} m \omega_x^2 x^2 ,
\end{equation}
where $V_0$ is the potential at the saddle and the curvatures of
the potential are parametrized by $\omega_x$ and $\omega_y$.
The resulting transmission probabilities
have the form of Fermi functions  
$T(E) = 1/(e^{\beta (E-\mu)}+ 1)$ 
(with a negative temperature 
$\beta = - 2 \pi /\hbar \omega_x$
and 
$\mu = \hbar \omega_y (n+1/2) + V_{0})$.
As a function of energy (gate voltage) the conductance rises 
step-like. 
The energy derivative of 
the transmission probability 
$dT_{n}/dE =  (2 \pi /\hbar \omega_x) T_{n}(1-T_{n})$
is itself proportional to the transmission probability times 
the reflection probability. We note that such a relation 
holds not only for the saddle point model of a QPC but also for 
instance for the adiabatic model \cite{glaz}. 
As a consequence 
$({1}/{4T_{n}R_{n}}) (dT_{n}/dE)^{2} =$
$({\pi}/{\hbar \omega_x})^{2} T_{n}R_{n}$
is proportional to $T_{n}R_{n}$. 
Thus the charge relaxation resistance of a saddle 
QPC is  
\begin{equation}
R_{q} = \frac{h}{4e^{2}} 
\frac{\sum_{n} \left[(\frac{d\phi_{n}}{dE})^{2} +  
(\frac{\pi}{\hbar \omega_x})^{2} T_{n}R_{n}\right]}
{[\sum_{n} \frac{d\phi_{n}}{dE}]^{2}} . 
\label{rqsad}
\end{equation}
\begin{figure}
\begin{center}
\includegraphics[scale=0.6]{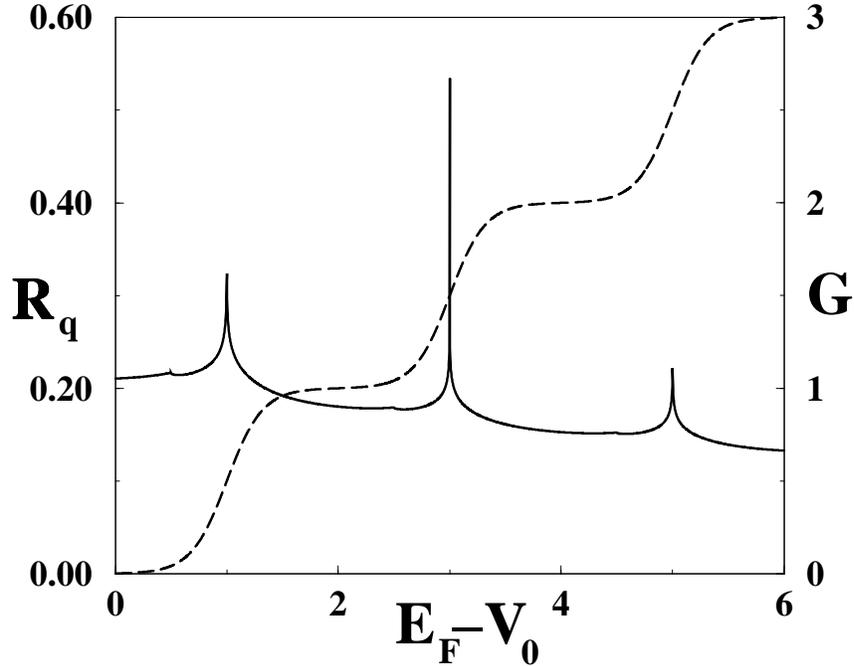}
\end{center}
\caption{Charge relaxation 
resistance $R_{q}$ of a saddle QPC in units of $h/4e^{2}$ for 
$\omega_y/\omega_x = 2$ and a screening length of 
$m\omega_x \lambda^{2} /\hbar = 2E_{\lambda}/\hbar \omega_{x} = 25$ 
as a function of $E_F -V_{0}$
in units of $\hbar \omega_x$ (full line). 
The broken line shows the conductance of the QPC. 
After \protect\cite{mbam} .}
\label{rqfig}
\end{figure}
To find the density of states 
of the $n$-th eigen channel, we use the relation between 
density and phase (action) and $D_{n} = \sum_{i} {\cal D}_{n,ii}
= (1/\pi) d\phi_{n}/dE$. 
We evaluate the phase semi-classically.
The spatial region of interest for which we have to find 
the density of states is the region over which the electron
density in the contact is not screened completely. 
We denote this length by $\lambda$ and the associated energy by
$E_{\lambda} = (1/2) m\omega_{x}^{2} \lambda^{2}$.  
The density of states is then found from  
$D_{n}=1/h \int_{-\lambda}^{\lambda} \frac{dp_n}{dE} dx$ 
where $p_{n}$ is the classically allowed momentum.
A calculation gives a density of states \cite{plb}
$D_{n} (E) = ({4}/({h\omega_x}))$
${\rm asinh}[E_{\lambda}/({E-E_n})]^{1/2}$, 
for energies $E$ exceeding the channel threshold $E_n$ and gives 
a density of states 
$D_{n} (E) = ({4}/({h\omega_x}))$
${\rm acosh}[E_{\lambda}/({E_n -E})]^{1/2}$
for energies in the 
interval $E_n - E_{\lambda}$
$ \leq E < E_n$ 
below the channel threshold. 
Electrons with energies less than $E_n - E_{\lambda}$
are reflected before
reaching the region of interest, 
and thus do not contribute to the density of states.
The resulting density of states has a logarithmic singularity
at the threshold $E_{n}= \hbar\omega_y(n+\frac{1}{2}) + V_0$  
of the $n$-th quantum channel. (A fully quantum mechanical 
calculation gives a density of states which exhibits also
a peak at the threshold but which is not singular). 

We now have all the elements to calculate the charge relaxation 
resistance $R_q$ and the resistance $R_v$. 
The charge relaxation 
resistance for a saddle QPC is shown in Fig. \ref{rqfig} for a set 
of parameters given in the figure caption. The charge relaxation 
resistance exhibits a sharp spike at each opening of a quantum channel. 
Physically this implies that the relaxation of charge, determined by 
the $RC$-time is very rapid at the opening of a quantum channel.

\begin{figure}
\begin{center}
\includegraphics[scale=0.6]{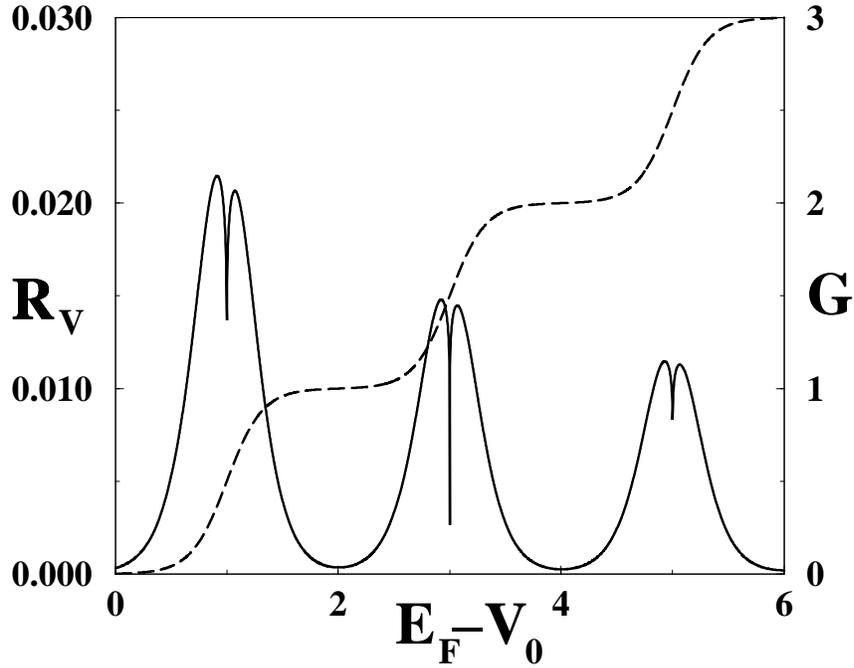}
\end{center}
\caption{$R_{v}$ (solid line) for a saddle QPC 
in units of ${h/e^{2}}$ and $G$ (dashed line)
in units of ${e^{2}/h}$ as a function of $E_{F} - V_{0}$ in units of 
$\hbar \omega_x$ 
with $\omega_y/\omega_x = 2$ 
and a screening length $m \omega_x \lambda^{2}/\hbar = 25$.
$R_v$ and $G$ are for spinless electrons. }
\label{rvedge}
\end{figure}

Similarly, we can find the resistance $R_v$ for a QPC
subject to a voltage $e|V|>>kT$.  
Using the density matrix elements 
for a symmetric QPC given by Eqs. (\ref{mqpc}),  
we find \cite{plb} 
\begin{eqnarray}
R_v =
\frac{h}{e^2} 
\frac{ \sum_n \frac{1}{4R_{n}T_{n}}
	\left( \frac{dT_{n}}{dE} \right)^{2}}
{[\sum_{n} (d\phi_{n}/dE)]^{2}}	
= {\frac{h}{e^{2}}} 
(\frac{\pi}{\hbar \omega_x})^{2}
\frac{\sum_n T_{n}R_{n}}{[\sum_{n} (d\phi_{n}/dE)]^{2}} .
\label{rvqpc1}
\end{eqnarray}	
The resistance $R_v$ is shown in Fig. \ref{rvedge}. 
\section{Dephasing rates and potential fluctuations} 
Let us now use the results discussed above to find the dephasing rates 
in the Coulomb coupled conductors. 
Consider a scattering state $\Psi_{i}({\bf r},E)$ at energy $E$ 
in conductor $i$ 
which solves the Schr\"{o}dinger equation for a fixed potential 
$U_{eq,i}({\bf r})$. Fluctuations of the potential 
$U_{i} ({\bf r},t)$
away from the static (average) equilibrium potential
will scatter the carrier out of the eigenstate $\Psi_{i}({\bf r},E)$.
Here we regard the fluctuating potential in the interior of the conductor 
as spatially uniform. 
Thus the fluctuating 
potential $dU_{i} (t) = U_{i}(t) - U_{eq,i}$
in the region of interest is a function of 
time only. The effect of the fluctuating potential 
can then be described with the help of a time-dependent phase 
$\phi (t)$ which multiplies the scattering state. Thus we consider 
a solution of the type 
$\Psi_{i}({\bf r},E) {\rm exp}(-i\phi (t))$ 
of the time dependent Schr\"odinger equation. 
The equation of motion for the phase is simply, 
$\hbar d\phi/dt = e dU_{i}(t)$.
Let us characterize the potential fluctuations in conductor $i$
by its noise spectrum $S_{U_{i}U_{i}}(\omega)$. Using 
the noise spectrum $S_{U_{i}U_{i}}(\omega)$
of the voltage fluctuations
we find that at long times the 
phase $\phi$ of the scattering state diffuses with a rate 
\begin{equation}
\Gamma^{(i)}_{\phi} = {\langle (\phi(t) - \phi(0))^{2} \rangle}/{2t} =
({e^{2}}/{2\hbar^{2}}) S_{U_{i}U_{i}}(0)
\label{dr1}
\end{equation} 
determined by the zero-frequency limit of the noise power 
spectrum of the potential fluctuations.

Eq. (\ref{ub}) shows that the potential fluctuations 
and thus the dephasing rate has two sources: A carrier in conductor 
$i$ suffers dephasing due to charge fluctuations in conductor $j =i$ itself, 
and due to charge fluctuations of the additional nearby conductor 
$j \ne i$. Accordingly, we can also write the dephasing rate
in conductor $i$ as a sum of two contributions, 
$\Gamma^{(i)}_{\phi} = \sum_{j} \Gamma^{(ij)}_{\phi}$ 
with 
\begin{equation}
\Gamma^{(ij)}_{\phi} = (e^{4}/2\hbar^{2}) G^{2}_{ij} S_{N_{j}N_{j}}(0) .
\label{dephgn}
\end{equation} 
At equilibrium, we can express the charge noise power with the help 
of the equilibrium charge relaxation resistances $R^{j}_q$ given by
Eq. (\ref{rq}). Of particular interest is the dephasing rate 
in the cavity (conductor $i = 0$) due to the presence of 
the QPC (conductor $i =1$). The self-consistent theory 
gives for this dephasing rates at equilibrium \cite{mbam} 
\begin{equation}
\Gamma^{(01)}_{\phi} = (e^{2}/\hbar^{2}) (e^{2}D_{1}G_{01})^{2} R^{(1)}_{q} kT .
\label{g12rq}
\end{equation}
In the non-equilibrium case,
if conductor $1$ is subject to a current generated by a voltage $|V|$
(with $kT <<e|V|$) we find a dephasing rate \cite{mbam} 
\begin{equation}
\Gamma^{(01)}_{\phi} = (e^{2}/\hbar^{2}) 
(e^{2}D_{1}G_{01})^{2} R^{(1)}_{v} e|V|.
\label{g12rv}
\end{equation}
These dephasing rates are proportional to $(e^{2}D_{1}G_{01})^{2}$.
Thus $e^{2}D_{1}G_{01}$ plays the role of an effective coupling 
constant in our problem. 
In the limit of a small Coulomb energy
($C \gg e^{2}D_{0}$ and $C \gg e^{2}D_{1}$)
we find $e^{2}D_{1}G_{01} = D_{0}/(D_{0}+ D_{1})$
{\it independent} of the geometrical capacitance $C$. 
In the limit of a large Coulomb energy 
($C \ll e^{2}D_{0}$ and $C \ll e^{2}D_{1}$) the effective coupling constant
becomes proportional to the capacitance 
$e^{2}D_{1} G_{01} = C/(e^{2} D_{0})$. The second limit, typically, 
is the physically relevant limit. 

According 
to Eq. (\ref{g12rq}) and Eq. (\ref{g12rv}), 
in an experiment in which the QPC is opened
with the help of a gate, the dephasing rate follows, at equilibrium
just $R_q$ and in the non-equilibrium case
follows $R_v$. 
Without screening $R_v$ would exhibit a bell shaped behavior as a function 
of energy, i. e. it would be proportional to $T_{n}(1-T_{n})$
in the energy range in which the n-th transmission channel is 
partially open. Screening, which in $R_v$ is inversely proportional to the 
density of states squared, generates the dip at the threshold 
of the new quantum channel at the energy which corresponds to $T_{n} = 1/2$
(see Fig. \ref{rvedge}). 
It is interesting to note that the experiment \cite{buks1} does 
indeed show a double hump behavior of the dephasing rate.

\section{Self-consistent measurement time}

Suppose that a measurement of the current is used to 
determine the charging state of the cavity. 
Consider the two charging states 
with $Q_0$ and $Q_0 +e$ electrons on the cavity. 
The two charge states on the cavity 
give, via long range Coulomb
interaction, rise to a conductance $G$ (charge $Q_0$)
and $G + \Delta G$ (charge $Q_0+e$). The measurement time 
$\tau_m$ is the (minimal) time needed to determine 
the conductance through a measurement that flows
through the QPC. 
A measurement needs to overcome 
the fluctuations of the current 
(shot noise and thermal noise). 
The measurements needs to be 
long enough \cite{aleiner1,buks1} so 
that the integrated current $\Delta G |V| \tau_m$ 
due to the variation of the conductance,
exceeds the integrated current fluctuation
$ \sqrt{S_{II}(0) \tau_m}$. Here,
$S_{II}(0)$ is the low frequency spectral current density,  
which in the zero temperature limit is due to shot nose 
alone and given by \cite{leso,mb92,physr}
$S_{II}(0) = 2e (e^{2}/h) |V| \sum_{n} T_{n}R_{n}$. 
This gives a measurement time 
\begin{equation}
\tau_m = \frac{S_{II}}{(\Delta G)^{2} |V|^{2}} . 
\label{taum}
\end{equation} 
$\Delta G$ is another quantity that can be measured independently
and such a measurement has in fact been carried out 
by Buks et al. \cite{buks1}. 
It is thus useful to make a theoretical prediction for this 
quantity. Therefore our purpose here is to evaluate $\Delta G$
and compare the expression for the measurement time 
with the phase breaking rate obtained above. 

The variation of the conductance of the QPC is 
determined by the sensitivity of the conductance due 
to the variation of the 
potential $dU_{1}$ in the QPC
\begin{equation}
\Delta G = (e^{2}/h) (dT/dU_{1}) dU_{1} . 
\label{dg1}
\end{equation} 
Here $T$ is the total transmission probability, 
$T \equiv \sum_{n} T_n$. In Eq. (\ref{dg1}) $U_{1}$ is the potential 
in the QPC and 
$dU_{1}$ is the change in potential for the case that an additional 
electron eners the cavity. 
In WKB-approximation we are allowed to replace the derivative with respect 
to the potential with a derivative with respect to energy, 
$dT/dU_{1} = - e dT/dE$. For the saddle point QPC we 
have $dT/dE = (2\pi/\hbar \omega_x) \sum_n T_n R_n$. Form Eq. (\ref{q5}) we find with 
$e dN_{0} = e$,  
a potential variation $dU_{1} = G_{10} edN_{0} = e G_{10}$ where 
$G_{10}$ is an off-diagonal element of the effective interaction 
matrix, Eq. (\ref{geff}). Thus the addition of an electron onto the cavity 
changes the conductance by \cite{br2} 
\begin{equation}
\Delta G = - (e^{2}/h) eG_{10} dT/dE .
\label{dg2}
\end{equation} 
Since $G_{10} = C_{\mu}/(e^{2}D_0 e^{2}D_1)$ 
we find in the limit $e^{2}/C >> 1/D_1$, $e^{2}/C >> 1/D_2$,  
\begin{equation}
\Delta G = - (e^{2}/h) (C/e^{2}D_0 ) (e (dT/dE)/e^{2}D_1) .
\label{dg3}
\end{equation} 
As discussed above, as a function of gate voltage the density of 
states $D_1$ of the QPC exhibits a strong 
variation. In particular, at zero temperature, the density 
of states diverges at the threshold of a new quantum channel. 
Consequently, $\Delta G$ is also a strong function of gate voltage. 
$\Delta G$ vanishes on the conductance plateaus since $(dT/dE)$
vanishes on a plateau. Eq. (\ref{dg2}) predicts that 
$\Delta G$ vanishes also at the channel opening threshold ($T_{n} =1/2$)
since the semiclassical density of states diverges. 
Thus Eq. (\ref{dg2}) predicts that $\Delta G$
(like $R_v$) is maximal away from the channel opening 
threshold. 
Such a behavior is not seen in the experiment 
of Buks et al. \cite{buks1}, $\Delta G$ seems to be rather independent 
of the gate voltage. This can be due to the simple model of the QPC 
used here or due to the fact that the experiment is not in the 
zero-temperature limit.

We now return to the measurement time, Eq. (\ref{taum}). 
Using Eq. (\ref{dg2}) and $dT/dE = (2\pi/\hbar \omega_x) \sum_n T_n R_n$, 
and the expression for $R_v$ as given by Eq. (\ref{rvqpc1}) 
we find,

\begin{equation}
\tau_m = 
\frac{2e^{2} }
{\pi^{2}(e^{2}/h)^{2}[e^{2} G_{10} D_{1}]^{2} R_v e|V| } .
\label{taum2}
\end{equation}

Comparison with our result for $\Gamma_{\phi}$ (see Eq. (\ref{g12rv}))
shows that $(1/2) \Gamma_{\phi} \tau_m = 1$ which agress with 
Korotkov \cite{korot}. As mentioned in the introduction, 
the identification of the measurement time $\tau_m$ with $1/\Gamma_{\phi}$
is taken for granted by a several authors. 
It is now, however, clear 
that in general \cite{ss,korot}
$(1/2) \Gamma_{\phi} \tau_m > 1$ 
( for instance for a detector that is not symmetric). 
Even for the symmetric detector (symmetric QPC) considered here, it is clear
that a non-zero temperature has a different effect on the dephasing time 
and on the measurement time. The dephasing time $\tau_{\phi} = 1/\Gamma_{\phi}$
is inversely proportional to the charge fluctuation spectrum 
whereas the measurement time is proportional to the current fluctuation
spectrum. 
At elevated temperatures, 
the measurement must overcome the combined thermal and shot noise 
and the measurement time will thus increase 
with increasing temperature. On the other hand 
the additional Nyquist noise leads to a shorter dephasing time.

\section{Discussion}

We have presented an electrically
self-consistent discussion of dephasing rates 
and measurement times for 
Coulomb coupled conductors. The approach emphasizes that also in 
open conductors, like QPC's, charge fluctuations are associated 
with a Coulomb energy. In such a self-consistent treatment, the dephasing 
rates are typically not simply proportional to a coupling constant.
In this work, we have  
attributed only a single potential 
to each conductor, but the theory \cite{mbam} 
is not in fact limited to such a
simplification and permits the treatment of 
an arbitrary potential landscape \cite{mbam,ammb}.
The theory also permits a discussion of a wide variety of 
geometries \cite{ammb,ankara}. 

We have treated the charge fluctuations within a linear screening approach. 
Large changes in the potential of the QPC would 
require a discussion of non-linear screening. In either case, a theory is 
necessary which treats the true charge distribution
and its fluctuations. A carrier on the 
cavity is entangled not only with a single electron on the QPC but 
with all electrons which are involved in the screening process on the cavity 
and on the QPC. Instead of a few electron problem our 
approach emphasizes the true many body nature of charge fluctuations of 
Coulomb coupled conductors. We have restricted our considerations to the 
case where we deal with open conductors for which charge quantization 
plays a minor role. The considerations given above apply, however, also 
to the case where we have a QPC interacting with a system in which charge is 
quantized. Even in this case carriers on the QPC will be screened to a certain 
extent and the charge relaxation resistance $R_q$ and the resistance $R_v$ 
should again be part of a self-consistent answer.

\section*{Acknowledgement}

This work was supported by the Swiss National Science Foundation and 
the TMR network.

%


\begin{thebibliography}{10}
\bibitem{note1}    The Lindhard function is 
                   $\pi({\bf r}, {\bf r^{\prime}}) \equiv \partial 
                   \rho({\bf r})/\partial U({\bf r^{\prime}})$ with $\rho$
                   the local density and $U$ 
                   the local potential. Since we assume a single spatially 
                   uniform potential and 
                   since we are interested in the total charge on the 
                   conductor in the volume (area) $\Omega$ we have 
                   $\int d^{3}{\bf r} \int d^{3}{\bf r^{\prime}} 
                   \pi({\bf r}, {\bf r^{\prime}}) = - e^{2} 
                   \int d^{3}{\bf r} \nu({\bf r}) = - e^{2}D$. 
                   Here $\nu(\bf{r})$ is the local density of states 
                   and $D$ denotes the total integrated density 
                   of states of the conductor.
\end{thebibliography}
\end{document}